\documentclass[twocolumn,secnumarabic,amssymb, showpacs, nobibnotes, aps, prd]{revtex4-1}

\usepackage{graphicx}
\usepackage{subfigure}
\usepackage{ae} %pdf-Schriftarten
\usepackage{float}  % allows to use the [H] parameter for includegraphics- placed the picture exactly at this place.
\usepackage{amsmath}
\usepackage{color}
\usepackage{siunitx}

\begin{document}

\title{Anatomy of ultrafast quantitative magneto-acoustics in freestanding nickel thin films}

\author{Antonia Ghita$^{1}$}
\author{Tudor-Gabriel Mocioi$^{1}$}
\author{Alexey M.~ Lomonosov$^{2}$}
\author{Jiwan Kim$^{3}$}
\author{Oleksandr Kovalenko$^{1}$}
\author{Paolo Vavassori$^{4,5}$}
\author{Vasily V.~ Temnov$^{1}$}
\email[]{vasily.temnov@cnrs.fr}

\affiliation{$^1$LSI, Ecole Polytechnique, CEA/DRF/IRAMIS, CNRS, Institut Polytechnique de Paris, F-91128, Palaiseau, France}
\affiliation{$^2$B+W Department, Offenburg University of Applied Sciences, 77652 Offenburg, Germany}
\affiliation{$^3$Department of Physics, Kunsan National University, 54150 Kunsan, Korea} 
\affiliation{$^4$CIC nanoGUNE BRTA, E-20018 Donostia-San Sebastian, Spain} 
\affiliation{$^5$IKERBASQUE, Basque Foundation for Science, E-48013 Bilbao, Spain}
%\affiliation{$^2$Scientific and Technological Center of Unique Instrumentation, Russian Academy of Sciences, 117342, Butlerova str. 15, Moscow, Russian Federation}
%\affiliation{$^3$Syktyvkar State University named after Pitirim Sorokin, 167001, Syktyvkar, Russia}
%\affiliation{$^4$Chelyabinsk State University, 454001 Chelyabinsk, Russia}
%\affiliation{$^5$South Ural State University (National Research University), 454080 Chelyabinsk, Russia}

%\email[]{ngoc\_minh.tran@univ-lemans.fr}

\date{\today}

\begin{abstract}
 We revisit the quantitative analysis of the ultrafast magneto-acoustic experiment in a freestanding nickel thin film by Kim and Bigot \cite{kim2017magnetization} by applying our recently proposed approach of magnetic and acoustic eigenmodes decomposition by Vernik et al. \cite{vernik2022resonant}. We show that the application of our modeling to the analysis of time-resolved reflectivity measurements allows for the  determination of amplitudes and lifetimes of standing perpendicular acoustic phonon resonances with unprecedented accuracy. The acoustic damping is found to scale as $\propto\omega^2$ for frequencies up to 80~GHz and the peak amplitudes reach $10^{-3}$. The experimentally measured magnetization dynamics for different orientations of an external magnetic field agrees well with numerical solutions of magneto-elastically driven magnon harmonic oscillators. Symmetry-based selection rules for magnon-phonon interactions predicted by our modeling approach allow for the unambiguous discrimination between spatially uniform and non-uniform modes, as confirmed by comparing the resonantly enhanced magneto-elastic dynamics simultaneously measured on opposite sides of the film. Moreover, the separation of time scales for (early) rising and (late) decreasing precession amplitudes provide access to magnetic (Gilbert) and acoustic damping parameters in a single measurement.   
\end{abstract}

\pacs{Valid PACS appear here}

\keywords{ultrafast magnetization dynamics, magneto-acoustics, exchange magnons, nanomagnetism}

\maketitle
%\tableofcontents

\section{Introduction}
Since early experimental studies \cite{Scherbakov2010,kim2012ultrafast,Thevenard2010}, ultrafast magneto-elastic interactions driven by femtosecond light pulses are conveniently described in the time-domain: the dynamics of magnetization driven by single or multiple acoustic pulses with picosecond duration are monitored using the magneto-optical pump-probe technique.  This intuitive picture allows for an elegant description of magnetization precession amplified by a sequence of acoustic pulses with an appropriate time interval via the magneto-acoustic coherent control mechanism \cite{kim2015controlling,kim2017magnetization}. Moreover, the time-domain picture of ultrafast magneto-acoustics facilitates the interpretation of magnetization switching \cite{KovalenkoPRL2013,Vlasov2020}, where the duration of acoustic pulses is shorter than the period of ferromagnetic resonance (FMR) precession.         

An alternative view on ultrafast magneto-acoustics is provided in magneto-optical transient grating experiments \cite{janusonis2016ultrafast,Janusonis2016_1,chang2017parametric,chang2018driving}. Here, the spectrally separated quasi-monochromatic acoustic excitations allow for observing resonant amplification of FMR precession induced by each acoustic mode. The dependence of FMR frequency on the external magnetic field makes it possible to tune the FMR precession in resonance with an acoustic mode of interest. Very recently we have extended this approach to ultrafast magneto-acoustic dynamics in free-standing thin films and multilayers \cite{vernik2022resonant}. Our theoretical approach is based on eigenmodes decomposition of both acoustic and magnetization dynamics, which allows for a more insightful analysis of ultrafast magneto-acoustic dynamics experiments in terms of resonant magneto-elastic interactions between individual modes of longitudinal acoustic phonons and perpendicular standing spin wave (magnon) modes \cite{van2002all}. For instance, the application of such rigorous theoretical analysis to resonant phonon-magnon interactions in freestanding multilayer structures predicts the key role of the symmetry of magnetic and acoustic modes in prescribing well-defined selection rules for individual phonon-magnon interactions. One of the most relevant conclusions was that in symmetric structures interactions between magnon and phonon eigenmodes with different symmetries were forbidden.

In this paper, with the purpose of bench-marking the power of our improved approach, we apply it to reinterpreting the experimental results by Kim and Bigot \cite{kim2017magnetization} obtained for a 300~nm freestanding nickel thin film. We show that even for such thick structures, where frequencies of spatially uniform (FMR) and non-uniform (spin wave or magnon) modes cannot be distinguished using conventional approaches employed so far, our approach enables the detection of their excitation thanks to the symmetry-dependent selection rules that govern their resonant interaction with acoustic modes. The results of this work are multiple: from the one side they demonstrate the ability of our modeling to retrieve fundamental parameters governing the complex physics involved in ultrafast magneto-acoustic experiments with an unprecedented accuracy, from the other side corroborate that the physical picture embodied in our model is particularly insightful, for example by highlighting the importance of symmetries in magneto-acoustics.

\section{Experiment}

Freestanding nickel membranes in the experiment in Ref.~\cite{kim2017magnetization} had a thickness $L$=300~nm and were obtained by depositing Ni on a glass substrate with a layer of sodium chloride in between them. The layer was subsequently dissolved in water to leave the Ni film stretched on a sample holder with a hole. The film was stretched laterally by gluing a silver paste around the edges of the film, which created a static strain in Nickel upon drying out.
\begin{figure}
	\footnotesize{} \centering
 \includegraphics[width=1.0\columnwidth]{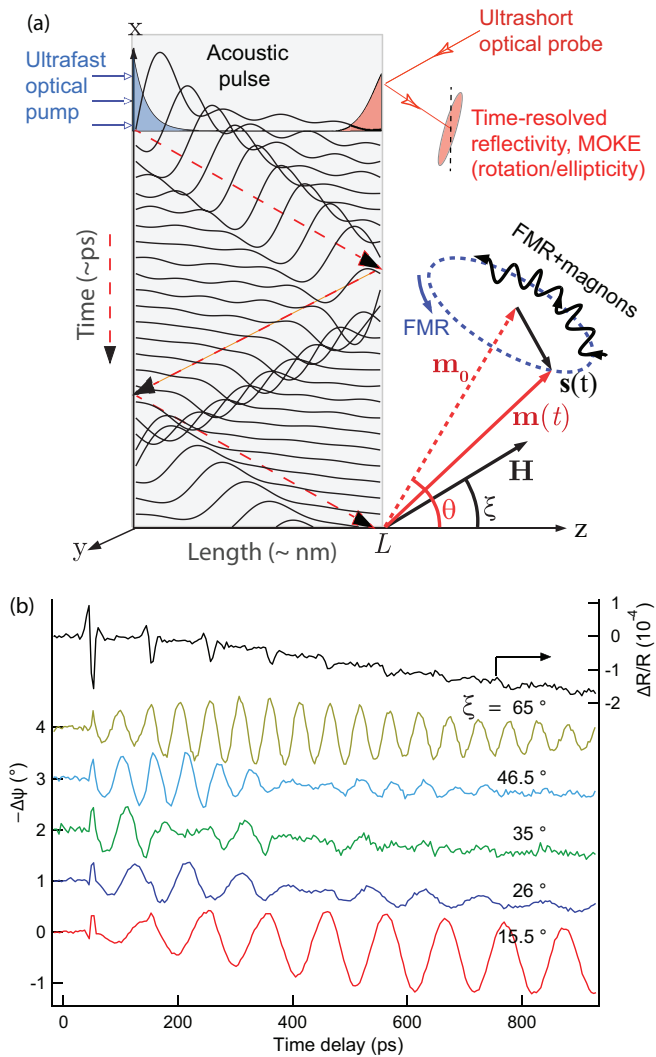}
 \caption{\label{Fig1} (a) Schematic picture of the experiment and acoustic pulse propagating inside the sample. The shaded exponentially decaying functions illustrate the optical penetration depth of pump and probe pulses, respectively.  (b) Experimental data for reflectivity and Kerr rotation.}
\end{figure}

The nickel thin film was optically excited at the front side by a femtosecond pump pulse (400~nm wavelength, 45~fs pulse duration, 10~kHz repetition rate, 1.5~mJ/cm$^2$ fluence), launching pulses of coherent longitudinal acoustic phonons with a duration of a few picoseconds propagating inside the sample at a constant speed $c_s$=6~nm/ps (see Fig.~\ref{Fig1}(a)). Due to the inverse magnetostrictive effect, these acoustic pulses drove the magnetization dynamics inside the Ni film. Time-delayed probe pulses of $\SI{800}{\nm}$ detected transient changes in the reflectivity and magneto-optical Kerr effect (MOKE) rotation both at the front and back sides of the sample. A rotating permanent magnet positioned on top of the sample produced a magnetic field with reported magnitude $\mu_0 H\sim \SI{0.4}{\tesla}$ at a variable angle $\xi$ with respect to the surface normal. Due to the magnetic anisotropy the equilibrium direction of magnetization was non-collinear with the external magnetic field and made an angle $\theta$ with the surface normal.

Fig.~\ref{Fig1}(b) shows the measured differential reflectivity $\frac{\Delta R}{R}$ and Kerr rotation $\psi$ at the back side of the film for five different orientations of the external magnetic field: $\xi = {15.5}^\circ$, ${26}^\circ$, ${35}^\circ$, ${46.5}^\circ$ and ${65}^\circ$. The slowly varying thermal background in reflectivity and Kerr rotation signals originated from heat diffusion from the front to the back side of the film and will be subtracted throughout the manuscript in order to facilitate the quantitative comparison with simulations of rapidly varying elastic and magneto-elastic transients. Complementary Kerr rotation and reflectivity measurements have been performed at the front side of the film: these data will be introduced and discussed in Fig.~2(a) for reflectivity and Fig.~4(a) for Kerr rotation.  

\section{Physical model}
Excitation of acoustic and magnetic transients in ferromagnetic nickel with femtosecond laser pulses can be adequately described by the phenomenological two-temperature model (TTM) \cite{anisimov1974electron}, which governs the phenomena of ultrafast demagnetization on a deeply sub-picosecond time scale \cite{beaurepaire1996ultrafast,gudde1999magnetization} and generation of ultrashort acoustic pulses on a picosecond time scale \cite{saito2003picosecond}. In the current paper, we are going to disregard the transient ultrafast demagnetization and focus on the interaction between fs-laser-generated acoustic pulses and magnetization dynamics. Within the framework of the TTM, the non-equilibrium hot electrons are initially generated through the absorption of an optical pump pulse within its skin depth. Subsequently, they transport energy into the depth of the sample via electron diffusion and heat up the cold lattice via electron-phonon scattering. These complex spatio-temporal dynamics result in the emission of picosecond acoustic pulses caused by the thermal expansion of rapidly heated lattice.  In case of a freestanding nickel film, these acoustic pulses generated at the front side of the sample, propagate through the film, are reflected at the back Ni/air interface (with a reflection coefficient equal to -1), and keep bouncing back and forth between these two interfaces before they decay due to various phonon scattering mechanisms.   

The magnetization dynamics induced by such ultrashort acoustic pulses can be adequately described by a phenomenological approach using magneto-elastically driven Landau-Lifshitz-Gilbert (LLG) equations \cite{Scherbakov2010,Thevenard2010,kim2012ultrafast}. Adapted to the experimental geometry in Fig.~1, 
the phenomenological free energy density $F = F_Z + F_d + F_{ex} + F_{me}$ takes into account the Zeeman term $F_Z = -\mu_0 M_0 {\bf m} \cdot {\bf H}$ due to the presence of the external magnetic field ${\bf H}$, the anisotropy energy $F_d = \left( \frac{1}{2} \mu_0 M_0^2 + K \right) m_z^2$ consisting of the thin-film shape anisotropy and the phenomenological anisotropy constant $K$ due to static built-in strains in a stretched nickel membrane, the exchange energy $F_{ex} = \frac{1}{2} M_0^2 \sum_{i=1}^3 D \left( \frac{\partial \bf m}{\partial x_i} \right)^2$ and the magneto-elastic energy $F_{me}(t) = b_1 m_z^2 \varepsilon_{zz} (z,t)$  ($b_1\simeq 10^7$J/m$^3$ for Nickel \cite{chikazumi1997physics}) due to the interaction with an acoustic pulse $\varepsilon_{zz}(z,t)$. The relation between the angle $\xi$ of the magnetic field and  $\theta$  magnetization at equilibrium is given by:
\begin{equation}
\sin(\theta-\xi)=\frac{\Tilde{M}}{2H}\sin 2\theta\,,
\label{fmr_equation for the angles}
\end{equation}
where $\Tilde{M} = M_0 + \frac{2K}{\mu_0 M_0}$ and $M_0$ is the saturation magnetization for Ni.
The length of the magnetization vector stays constant in our model (assuming constant temperature), so the magnetization dynamics can be described with the unit magnetization vector ${\bf m}$ and its precession ${\bf s}(z,t)$:
\begin{equation}\label{magnetization_precession}
    {\bf m} = {\bf m}_0 + {\bf s}(z,t)\,,
\end{equation}
which can be represented as a sum of magnetic eigenmodes:
\begin{equation}
    {\bf s}(z,t)=\sum_{n=0}^\infty {\bf s}^{(n)}(t)\cos\left(k_n z\right) \,.
\end{equation}
 where $k_n = \pi n/L$ is the wavevector of the $n$-th magnetic eigenmode and free boundary conditions for magnetization dynamics are assumed.
The $n=0$ magnetic eigenmode with uniform spatial profile corresponds to ferromagnetic resonance (FMR), while higher-order $n\geq 1$ modes describe spatially non-uniform modes of exchange magnons.

It has been shown \cite{vernik2022resonant,BesseJMMM} that in the linear approximation when the acoustic strains are small, the magneto-elastically driven dynamics for each magnon mode satisfy the equation of a damped driven harmonic oscillator,
\begin{equation}\label{magnon_oscillator}
\frac{d^2s_z^{(n)}}{dt^2}+2\alpha\omega_n\frac{ds_z^{(n)}}{dt} + \omega_n^2s^{(n)}_z = f_n(t)\,,
\end{equation}
where $\alpha$ is the Gilbert damping parameter and magnon eigenfrequencies $\omega_n$ obey
\begin{widetext}
\begin{equation}
\label{fmr_magnon_tilted}
\omega_n=\gamma\mu_{0}\sqrt{\left(H\cos\xi-
\left(\Tilde{M}-{\tilde D}k_n^2\right)\cos\theta\right)^2+\left(H\sin\xi+{\tilde D}k_n^2\sin\theta\right)
\left(H\sin\xi+\left(\Tilde{M}+{\tilde D}k_n^2\right)\sin\theta\right)}\,.
\end{equation}
\end{widetext}
Here ${\tilde D}=D/(\hbar\gamma\mu_0)$ is the exchange stiffness ($D=\SI{430}{meV} \text{\AA} ^2$ from Ref. ~\cite{van2002all}) and $\gamma$ denotes the gyromagnetic ratio.

\begin{figure*}
	\footnotesize{} \centering
 \includegraphics[width=2.0\columnwidth]{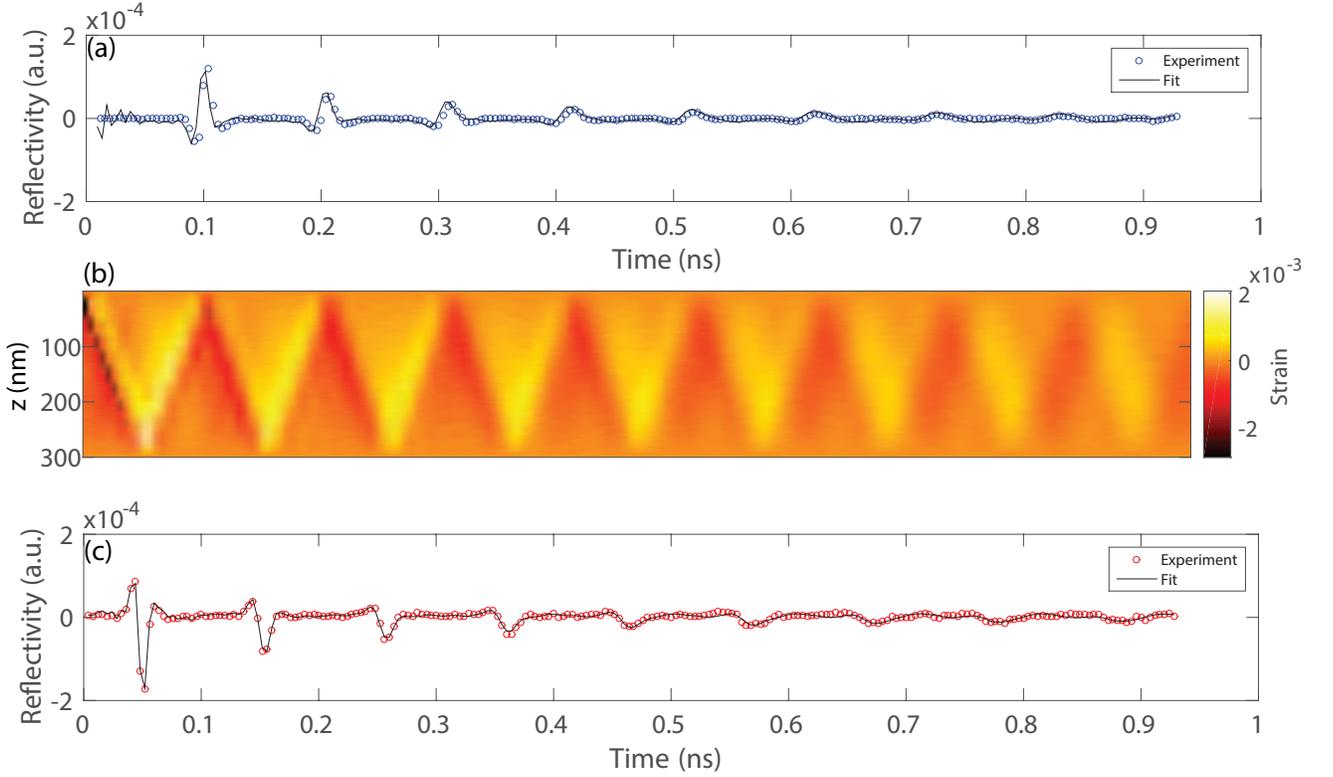}
 \caption{\label{Fig2} (a) Experimentally measured reflectivity at the front side of the Ni film, superimposed with its fit as a superposition of 10 decaying sinusoidal eigenmodes. (b) Color map showing the reconstituted strain inside the film as a function of position and time. (c) Experimental reflectivity measurement at the back side of the Ni film, together with its fit.}
\end{figure*}

The external magneto-elastic driving force
\begin{equation}\label{driving_force}
    f_n (t) = P_n({\bf H}) \int_0^L \varepsilon_{zz} (z, t) \cos \left(k_n z\right) dz ,\
\end{equation}
is proportional to the overlap integral between the magnon eigenmode with the acoustic strain pulse $\varepsilon_{zz}(z,t)$.  For our experimental geometry the prefactor 
\begin{equation}
    P_n({\bf H}) = \frac{\mu_0 \gamma^2 b_1 \sin(2\theta) \left(\tilde{D} k_n^2 \sin{\theta} + H \sin{\xi}\right)}{M_0 L}\,
\end{equation}
is proportional to the magnetostriction coefficient $b_1$ and depends both on the magnitude and orientation of an external magnetic field $\bf H$.  

Understanding the magneto-elastic dynamics governed by Eq.~(\ref{magnon_oscillator}) is facilitated by decomposing the acoustic strain pulse in its eigenmodes according to  
\begin{equation}\label{eigenmodes_strain}
    \varepsilon_{zz}(z,t)=\sum_{p=1}^\infty \varepsilon_{zz}^{(p)}(z) e^{-\gamma_p t} \cos{(\omega_p t + \varphi_p)}\,.
\end{equation}
We assume acoustic eigenmodes to oscillate at frequencies $\omega_p=c_sk_p$ and decay with damping constants $\gamma_p$;  $\varphi_p$ denote their initial phases. In a freestanding film, the acoustic eigenmodes obey the free boundary conditions for the acoustic displacement (corresponding to zero strains at both Ni/air interfaces) resulting in
\begin{equation}\label{strain_modes}
    \varepsilon_{zz}^{(p)}(z) = a_p \sin \left(k_p z\right)\,,
\end{equation}
where $k_p = \pi p/L$ is the wavevector of the $p$-th acoustic eigenmode. 
Using the decomposition of the acoustic strain in its respective eigenmodes, the expression of the magneto-elastic driving force becomes
\begin{equation} \label{driving_force_2}
    f_n (t) = P_n({\bf H}) \sum_{p=1}^{\infty} I_{np} a_p e^{-\gamma_p t} \cos (\omega_p t + \varphi_p)\,.
\end{equation}
Here we have introduced the overlap integral 
\begin{equation} \label{OverlapIntegral}
I_{np} = \int_0^L \cos \left(k_n z\right) \sin \left(k_p z\right)dz
\end{equation}
between the $n$-th magnetic and $p$-th acoustic eigenmodes. 

To sum up this section, after having quantified the acoustic strain and decomposed it in its eigenmodes, we can use equations (\ref{magnon_oscillator} - \ref{fmr_magnon_tilted}) and (\ref{driving_force_2} - \ref{OverlapIntegral}) to simulate the time evolution of the magnetization precession. To solve Eq.~(\ref{magnon_oscillator}) numerically, we use the 4th-order Runge-Kutta method.

\section{Analysis of acoustic data}

\begin{figure}
	\footnotesize{} \centering
 \includegraphics[width=1.0\columnwidth]{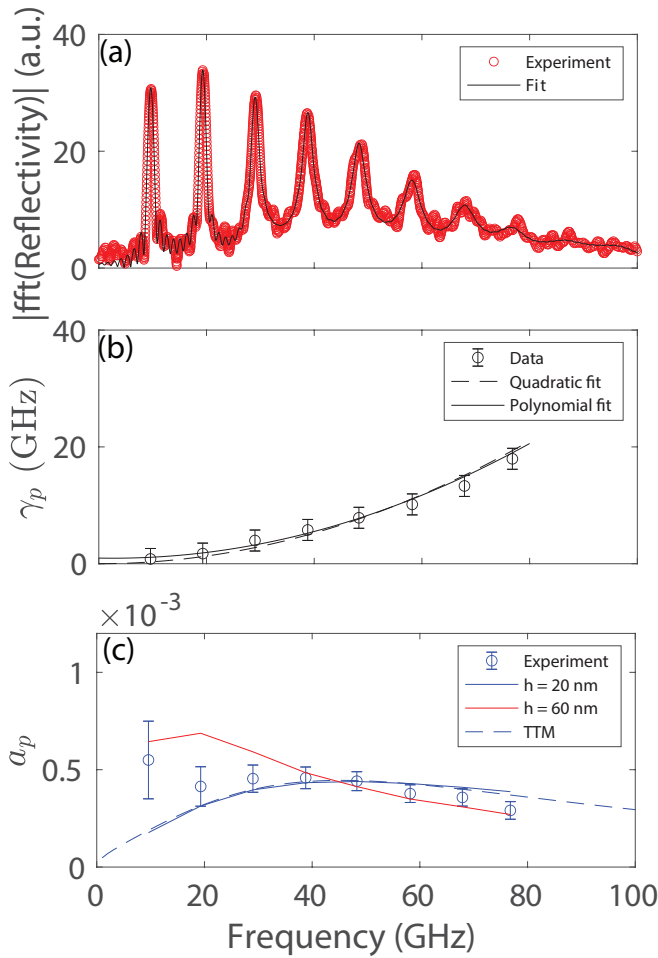}
 \caption{\label{Fig3} (a) Fourier Transform of the experimental reflectivity and its fit, for the backside of the film. (b) Damping as a function of frequency, as extracted from the fitting procedure. The result follows a quadratic law to a very good approximation. (c) Amplitudes of the acoustic modes, as extracted from the fitting procedure. The dashed line represents the acoustic frequency spectrum according to Eq.~(5) in Ref.~\cite{saito2003picosecond} for Nickel thin films excited by weak fs-pump pulses. , amplitudes obtained with Eq.~(\ref{a_p_simplified}) assuming exponential heating profiles with $h$=20~nm and 60~nm are shown for comparison (continuous lines).}.
\end{figure}

The change in reflectivity measured by the probe pulse after excitation with the pump pulse is related to the strain inside the film through the sensitivity function $f(z)$, which depends on the optical constants of Ni. Their relation (as in \cite{ThomsenPRB86}) is given by 
\begin{equation}\label{refl_strain}
    \frac{\Delta R(t)}{R} = 2\text{Re}\Big(\frac{\Delta r(t)}{r}\Big)=\int_{0}^L\varepsilon_{zz}(z,t) f(z) \mathrm{d}z\,,
\end{equation}
where we used the expression for the sensitivity function in the complex notation \cite{saito2003picosecond}:
\begin{equation}\label{senistivity_acoustic}
f(z) =  \frac{16 \pi}{\lambda} \text{Re}\left(i\frac{\partial\Tilde{n}}{\partial \varepsilon_{zz}}\frac{\Tilde{n}}{\Tilde{n}^2-1} e^{i \frac{4 \pi \Tilde{n}}{\lambda}z}\right)\,.
\end{equation}
Here $\Tilde{n}= 2.48 + 4.38 i$ denotes the complex index of refraction of Ni at the probe wavelength $\lambda = \SI{800}{\nm}$ and its derivative with respect to the applied strain   $\frac{\partial \Tilde{n}}{\partial \varepsilon_{zz}} = 0.6 - 1.8i$  \cite{saito2003picosecond} is called the photo-elastic coefficient. The spatial dependence of the sensitivity function is dominated by the exponential decay $\propto{\rm exp}(-z/\delta_{skin})$ within the optical penetration depth of the probe pulse $\delta_{skin} = \frac{\lambda}{4 \pi \text{Im}(\Tilde{n})} = \SI{14.5}{\nm}$.

Using the previous decomposition of strain into eigenmodes, we obtain the following expression for the measured reflectivity
\begin{equation}\label{refl_modes}
    \frac{\Delta R(t)}{R} = \sum_{p=1}^\infty a_p J_p e^{-\gamma_p t} \cos{(\omega_p t + \varphi_p)}\,,
\end{equation}
where we can interpret $J_p = \int_0^L f(z)\sin(k_p z)\mathrm{d}z\,$ as the detection integral of the $p$-th mode. This expression for the transient reflectivity shows that it can be represented as a sum of damped harmonic oscillations, suggesting that the Fourier transform of the signal could be useful in characterizing the acoustic eigenmodes. Panels (a) and (c) in Fig.~\ref{Fig2} show the reflectivity signal at the front and back side of the film, while Fig.~\ref{Fig3}(a) displays the Fourier transform of the backside reflectivity signal. The data allows us to distinguish ten peaks in the Fourier transform, so we carry out the analysis using the first ten acoustic modes. 

To obtain the amplitudes, lifetimes, and phases of the respective acoustic modes, we performed a non-linear least squares fitting (using the Levenberg-Marquardt algorithm) of the reflectivity data at the back side with Eq.~(\ref{refl_modes}), where  $a_p$, $\gamma_p$, and $\varphi_p$ are taken as fit parameters and frequencies $\omega_p$ are extracted from the Fourier spectrum. As an initial guess for the amplitudes $a_p$, we used the integrated intensity of each peak, and for damping constants $\gamma_p$ - the width of spectral lines from the Fourier spectrum. It is important to choose a rather precise initial guess for amplitudes and damping constants as the algorithm is highly sensitive to them. The results of such fitting in Fig.~\ref{Fig2}(a,c) and Fig.~\ref{Fig3}(a) appear to be in an excellent agreement with experimental data.

%Let us note that we used the reflectivity measurements at the back side to perform the fit since the pump pulse acted at the front side. While we can assume that the back side remained at room temperature, the temperature at the front side increased after heating with the pump pulse. In this case, the sensitivity function will change due to the variation of the optical constants of Ni with temperature. Thus, it is more justified to disregard the temperature dependence of the sensitivity function in equation \ref{senistivity_acoustic} when applying it to the data obtained at the back side, where temperature variations are significantly smaller. Kim: In the previous paragraph, it says that the back-side data was used. But in this sentence, Fig. 2 upper panel shows the front side data and its fit. Maybe switch Fig. 2(a) to Fig. 2(c)? Please check  We will need to change this wording. VT: Indeed, fits are there and they are not bad.

 Furthermore, panels (b) and (c) in Fig.~\ref{Fig3} show the dependence of damping and amplitudes for the first eight eigenmodes on their frequencies. The second-degree polynomial fit of damping $\gamma_p$ as a function of frequency shows that the quadratic term dominates. Thus, we can conclude that damping scales quadratically with frequency up to around $\SI{80}{\GHz}$. This result is consistent with Ref.~\cite{Maris}, suggesting that the attenuation mechanism is due to the phonon-phonon scattering. 
 
 The straightforward attempt to understand the fitted amplitudes (Fig.~\ref{Fig3}(c)) within the framework of the TTM failed. Using Eq.~(5) and the set of experimental fit parameters in the low-fluence excitation regime (pump fluence $\sim$0.01~mJ/cm$^2$) in Nickel thin films \cite{saito2003picosecond} results in the initial heat penetration depth $h$=20~nm that only slightly exceeds the optical skin depth of our pump pulses. In terms of the acoustic amplitudes the results of the TTM are well-approximated by a simplified phenomenological model assuming an instantaneous heating with an exponential profile $\propto\exp(-z/h)$ giving rise to    \begin{equation} \label{a_p_simplified}
a_{p}\propto \int_0^L e^{-z/h} \sin \left(k_p z\right)dz\,,
\end{equation} 
The strong disagreement between the theory and the experimental data indicates that this modeling cannot be applied. In the strong-excitation regime used in this experiment the parameters of the two-temperature model display strong dependence on the pump fluence \cite{lin2008electron} resulting in larger electronic heat capacity and weaker electron-phonon coupling. Both effects favor a larger heat penetration depth mediated by hot electron diffusion during the increase electron-phonon relaxation time. We can account for this effect by assuming a larger heat penetration depth $h$=60~nm, which provides a better approximation to the experimental data. However, it is clear that the discussed theoretical models represent oversimplifications and a further systematic study of the strong excitation regime of picosecond acoustic pulses is necessary.

Using the obtained amplitudes, phases, and damping parameters, we can reconstruct the strain inside the film as a function of space and time. Figure \ref{Fig2}(b) shows an evolution of spatial strain that is in accordance with the intuitive image of an acoustic echo propagating back and forth, undergoing reflections at both ends of the film and damping in time. But, in addition to this intuitive picture, the eigenmode decomposition also helps explaining the broadening of acoustic echo in time domain, which is due to the frequency-dependent damping.

\section{Analysis of magnetization dynamics}

\begin{figure}
	\footnotesize{} \centering
 \includegraphics[width=1.0\columnwidth]{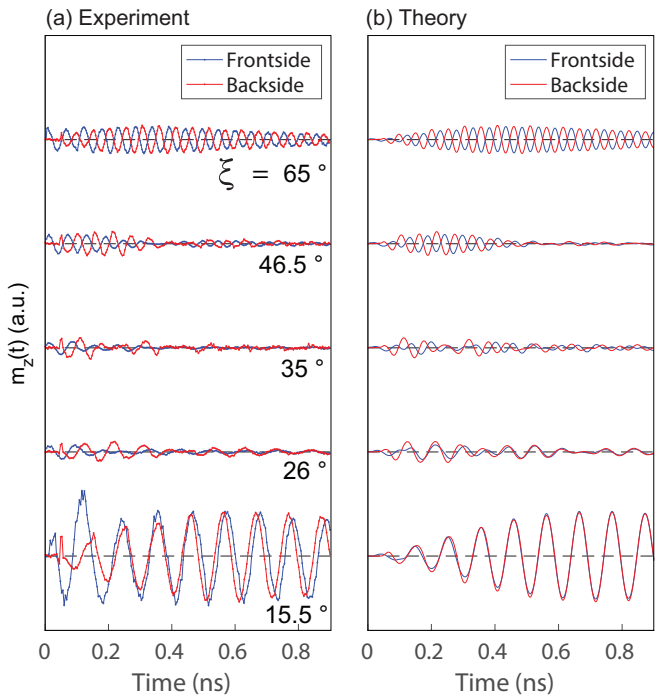}
 \caption{\label{Fig4} Comparison of magnetization dynamics at the front and back side of the film, as obtained (a) experimentally and (b) from our simulations.}
\end{figure}

Magnetization dynamics in the Ni film are analyzed by measuring the Kerr rotation angle. The depth sensitivity function of MOKE becomes important in case of ultrafast magnetization dynamics, varying within the skin depth of light due to the presence of spatially non-uniform high-frequency magnons. The relation between the detected Kerr rotation and the magnetization precession $s_z(z, t)$ inside the film is given by
\begin{equation}\label{MOKE_sensitivity}
\frac{\Delta \psi(t)}{\psi_s} = \int_0^L s_z(z, t)g(z) dz
\end{equation}
where $\psi_s$ is the static Kerr rotation angle, $\Delta\psi(t)$ is its change due to magnetization precession, and $g(z)$ is the depth sensitivity function for the polar MOKE \cite{traeger1992depthF}:
\begin{equation}
    g(z) = \frac{4 \pi}{\lambda} \text{Re}\Big (iQ_{MO}\frac{\Tilde{n}^2}{1+\Tilde{n}^2}e^{-i \frac{4 \pi \Tilde{n}}{\lambda}z}\Big ).
\end{equation}
Here, unlike the acoustical sensitivity function, the magneto-optical response is valued by the complex magneto-optical (Voigt) constant $Q_{MO}=i\frac{\Tilde{\epsilon}_{xy}}{\Tilde{\epsilon}_{xx}}=-(4.9+10.5 i)\times10^{-3}$ \cite{Krinchik1968}.

Using our previous decomposition in magnon eigenmodes, we get an expression that ties the dynamics of individual magnon modes to the Kerr rotation:
\begin{equation}\label{MOKE_sens_integral}
\frac{\Delta \psi (t)}{\psi_s} = \sum_{n=0}^{\infty} \Tilde{J_n} s_z^{(n)}(t)\,,
\end{equation}
where $\Tilde{J_n} = \int_{z=0}^L g(z) \cos \left(k_n z\right)dz$ is the detection integral of the $n$-th magnon mode. Using this expression, we can fit the results of our magnetization dynamics simulation with those of the experiment in the next section.

%Given this agreement, we can now look at the reconstructed magnetization precession profiles inside our sample (Fig.~5(b-c)). After the transient regime dies out, the magnetization profile for $15.5$ $^{\circ}$ seems to be more or less uniform in space, which suggests that the dynamics is dominated by the FMR ($n=0$) mode. On the other hand, the dynamics at $65$ $^{\circ}$ follow a spatially antisymmetric profile with respect to the middle of the film, which suggests that in this configuration the $n=1$ magnon dominates. This agrees with the phase shift between the data at the front and at the backside at $65$ $^{\circ}$, which we can notice in figure \ref{Fig4}. One of the intermediate angles, $35$ $^{\circ}$, shows a profile in space and time that is visually somewhat in between the two, however with a much lower amplitude. This suggests that in this intermediate regime, no magnon mode is resonantly excited.

\section{Resonant phonon-magnon interactions}

%Assuming a magnetic field of $\SI{0.30}{\tesla}$ and an anisotropy constant $K = 2.05 \cdot 10^5 \text{J/m}^{3}$, we arrive at a quantitative agreement between the experimental data and our simulations (figure \ref{Fig4}). The value of the magnetic field is in line with the typical value assumed in the experiment ($0.4$ T), but the large value and sign of $K$ can only be explained as originating from some static strain in the Ni film, i.e. an equilibrium value for $\varepsilon_{zz}(z,t)$ which is different from $0$. In our case, this residual strain would be equal to $2.05 \%$, which is plausible, given the specifics of the experiment. 

The experimental data for the back- and front side- Kerr rotation are presented in Fig.~\ref{Fig4}(a) for different orientations of the external magnetic field. We simulate the Kerr rotation by solving Eq.~(\ref{magnon_oscillator}) for each magnon and using the sensitivity function defined in the previous section to obtain the Kerr rotation from Eq.~(\ref{MOKE_sens_integral}). In order to reach an agreement with experimental data for all angles $\xi$ in Fig.~4, we have used the values for the anisotropy constant $K$, magnitude of the magnetic field $H$ and Gilbert damping $\alpha$ as fit parameters. Using the magnetic field of $\SI{0.3}{\tesla}$, the anisotropy constant $K = 2.05 \cdot 10^5 \text{J/m}^{3}$ and the Gilbert damping $\alpha = 0.04$, we achieve the quantitative agreement between the experimental data and simulations (Fig.~\ref{Fig4}). The value of the magnetic field stays within the expected error bar for a permanent magnet placed on top of the sample. The value of the Gilbert damping is equal to the one obtained in a recent study of ultrafast magnetization dynamics in nickel nanomagnets \cite{berk2019strongly}. The simulated Kerr rotation at the front and back sides are represented in Fig.~\ref{Fig4}(a,b). We observe an excellent agreement between the experimental data and simulations, except for the initial thermal excitation of magnetization at the front side, which we did not account for in our model. 
 
Given this agreement between experimental data and simulations, we analyze the peculiarities of the magnetization dynamics at different angles. For $15.5$ $^{\circ}$, the magnetization dynamics at the back side is in-phase with that at the front-side, while for $65$ $^{\circ}$ they are $\pi$-shifted. Moreover, we notice that the magnetization precession at $15.5$ and $65$ $^{\circ}$ lasts longer and is stronger than at other angles. While at $15.5$ $^{\circ}$ and $65$ $^{\circ}$ there are some slowly varying long-lived dynamics, at angles $26$, $35$, and $46.5$ $^{\circ}$ we observe weak, somewhat irregular beating patterns.

\begin{figure*} \label{Fig5}
	\footnotesize{}
 \includegraphics[width=1.8\columnwidth]{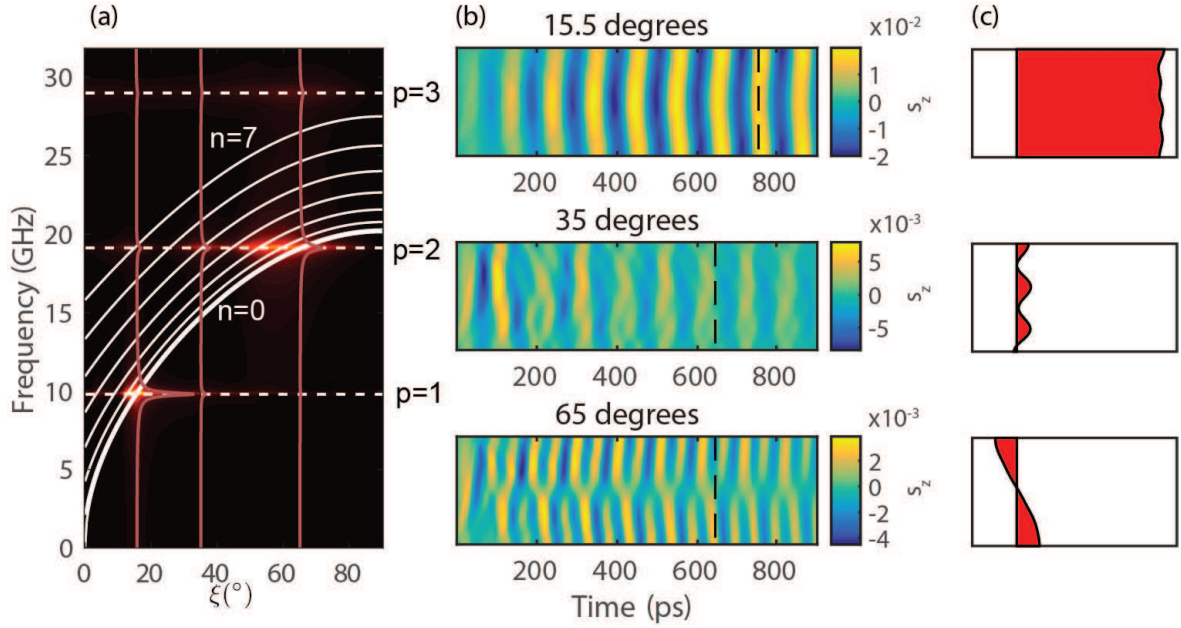}
 \caption{(a) Fourier transform of simulated Kerr rotation as a function of magnetic field angle. The dashed lines indicate the frequencies of phonons and the continuous white lines are the dispersion curves of magnons. Vertical cross-sections into the heat map, corresponding to the three angles in panel b, are shown. (b) Magnetization dynamics inside the film, as a function of position and time, for three experimental angles. (c) Magnetization profiles, taken at the times indicated in panel (b) by dashed lines.}
\end{figure*}
\begin{figure} \label{Fig6}
	\footnotesize{}
 \includegraphics[width=1.0\columnwidth]{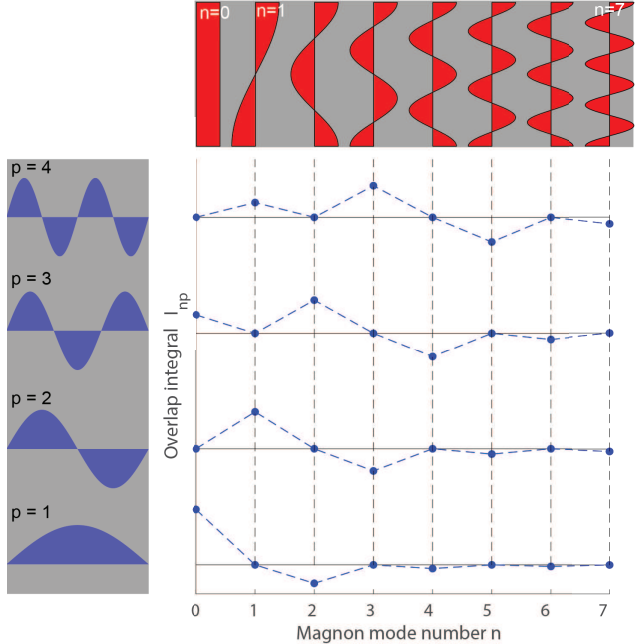}
 \caption{\label{overlap_int} Representation of magnonic eigenmodes (along the horizontal axis) and acoustic eigenmodes (along the  axis), with their corresponding overlap integrals.}
\end{figure}

A complementary perspective is presented in Fig.~5 which show the reconstructed magnetization dymnamics inside the sample (Fig.~5(b-c)) as a function of position and time. After the transient regime dies out, the magnetization profile for $15.5$ $^{\circ}$ is approximately uniform in space, suggesting that the dynamics is dominated by the FMR ($n=0$) mode. On the other hand, the dynamics at $65$ $^{\circ}$ follow the spatially antisymmetric profile with respect to the middle of the film, which suggests that in this configuration the $n=1$ magnon dominates. This conclusion is inline with the observed $\pi$-phase shift between the data at the front and the backside at $65$ $^{\circ}$, see Fig.~\ref{Fig4}. At an intermediate angle of $35$ $^{\circ}$, the magnetization dynamics with a much smaller amplitude are mainly visible at early delays times. This suggests that in this intermediate regime no magnon modes are resonantly excited.

All these observations can be explained by a simple theory for a driven harmonic oscillator. The main result is that the oscillation amplitude is resonantly enhanced when the natural frequency (here, that of magnons $\omega_n$) equals the driving frequency (in our case, that of phonons $\omega_p$). Away from resonance, the transient regime is characterized by a beating pattern because of the difference between the natural and driving frequencies.

Figure 5(a) shows the amplitude of magnetization precession as a function of its frequency and the angle of the external magnetic field. Two bright spots are visible: one at the point where the frequency of the first phonon ($p=1$) matches that of FMR ($n=0$) and another one where the frequency of the second phonon ($p=2$) matches that of the first magnon ($n=1$). 

However, for our symmetric freestanding membrane the overlap integral $I_{np}$ (Fig.~\ref{overlap_int}) is zero when the acoustic modes possess a different symmetry from that of the magnon modes. This means that symmetric (antisymmetric) acoustic modes will interact only with symmetric (antisymmetric) magnon modes, respectively. Therefore, the symmetry-based selection rules become as important for resonant phonon-magnon interaction as the previously mentionned frequency matching condition.

These considerations enable us to identify the driving forces of the magnetization dynamics observed at the three angles shown in Fig.~5. At $15.5^{\circ}$, the frequency of the first phonon ($p=1$) matches that of the first few magnons but the symmetry of the modes allows only even magnons to be excited. Since the overlap integral decays with increasing magnon number, the dominant magnetic mode at $15^{\circ}$ is the FMR ($n = 0$). At $65^{\circ}$, the second phonon ($p = 2$), whose frequency matches the frequencies of the first few magnons, interacts only with odd magnons. Thus, the dominant mode in this case is the first magnon ($n=1$). At $35^{\circ}$, the magnon frequencies are between the frequency of the first ($p=1$) and second ($p=2$) phonon and exhibit no resonant interaction. Hence, the magnetization precession is significantly weaker.

Figure 7 illustrates the quantitative agreement between theory and experiment for the two resonantly driven magnetization dynamics. There are two timescales involved in such phonon-magnon interactions: excitation (the transient regime) and decay (the driven regime). When the driving frequency is equal to the eigenfrequency, the response in the transient regime follows $(1 - e^{-t/\tau_{exc}})\cos{\omega_p t}$, where $\omega_p$ is the frequency of the acoustic driving mode and $\tau_{exc}$ is a characteristic relaxation time of magnetization dynamics, which is related to Gilbert damping as $\tau_{exc} =(\alpha \omega_n)^{-1}$, where $\omega_n$ is the frequency of the dominant magnon. On the other hand, the decay of magnetization is a driven regime when magnetization dynamics follow the acoustic driving force as $\propto e^{-t/\tau_{decay}}\cos{(\omega_p t)}$, where $\tau_{decay}=1/\gamma_p$ is the acoustic mode lifetime. Thus, to a good approximation, the overall magnetization dynamics fit inside an envelope of the form
\begin{equation}
    A(t) \propto (1 - e^{-t/\tau_{exc}}) e^{-t/\tau_{decay}}\,.
\end{equation}
 We can assign these characteristic relaxation times to the corresponding acoustic and magnetic eigenmodes, respectively, as shown in Fig.~7. For the magnetization dynamics at $15.5^{\circ}$, the characteristic time of the excitation phase is $\tau_{exc}^{(1)} = \SI{0.45}{\ns}$ and the decay time is $\tau_{decay}^{(1)} = \SI{1.2}{\ns}$. We can see a clear correlation between these two times and the corresponding lifetimes of the acoustic and magnetic eigenmodes, $\frac{1}{\gamma_{G}^{(0)}} = \SI{0.44}{\ns}$ and $\frac{1}{\gamma_{ac}^{(1)}} = \SI{1.19}{\ns}$. Similarly, for the case of magnetization dynamics at $65^{\circ}$, we extract the excitation time $\tau_{exc}^{(2)} = \SI{0.23}{\ns}$ and a value of the decay time $\tau_{decay}^{(2)} = \SI{0.57}{\ns}$. Again, this value is consistent with the lifetimes of the first magnon mode $\frac{1}{\gamma_{G}^{(1)}} = \SI{0.22}{\ns}$ and of the second acoustic eigenmode, $\frac{1}{\gamma_{ac}^{(2)}} = \SI{0.56}{\ns}$.

\begin{figure} \label{Fig7}
	\footnotesize{} 
 \includegraphics[width=1.0\columnwidth]{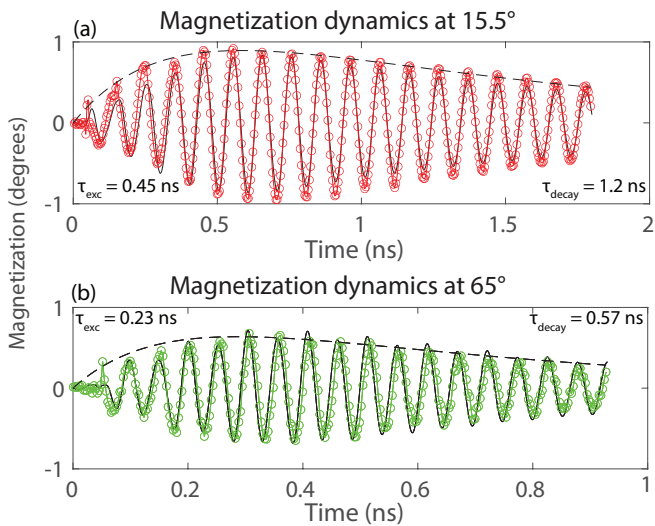}
 \caption{Long-scan magnetization dynamics at the backside of the film, under the magnetic field angles (a) $\xi=15.5^{\circ}$ and (b) $\xi=65^{\circ}$. Experimental data is represented by dots, the simulated curve is a black continuous line, while the dashed curve denotes the fit envelope $A(t)$. The separation of timescales is clearly visible in both graphs: the initial growth governed by the Gilbert damping is followed by the decay due to the acoustic decay.}
\end{figure}
This model also allows for extracting quality factors of magnons and phonon resonances. The magnon quality factor of $Q_{G} = 13\simeq1/(2\alpha)$ does not depend on the mode number, in agreement with previously reported results on frequency-independent Gilbert damping $\alpha$ \cite{razdolski2017nanoscale,Salikhov2019}. Acoustic quality factors $Q_{ac}^{(1)} = 37$ and $Q_{ac}^{(2)} = 29$ are slightly different due to the observed nonlinear dependence of acoustic damping on frequency, converging to the approximate scaling $Q_{ac}(\omega)\propto\omega^{-1}$ for higher-order acoustic modes. The observed high values of acoustic quality factors $Q_{ac}>Q_G$ enable the separation of time- scales in the excitation and decay phases in the magnetization dynamics. 

Qualitatively similar magnetization dynamics have been observed earlier in transient grating experiments \cite{janusonis2016ultrafast}. However, in the latter case the conspicuous decay of magnetization dynamics was explained by the complex spatio-temporal dynamics of the magnitude of the magnetization vector $M_z(x,t)$ on the temperature $T(x,t)$ in the periodically demagnetized nickel film \cite{Janusonis2016_1,chang2018driving}. In this context, our experimental geometry provides an advantage of isolating resonant phonon-magnon interactions from thermal effects and extracting their properties from the same measurement.  

\section{Summary and conclusions}
In this manuscript, we reported on the quantitative analysis of experimental data by Kim and Bigot in a free-standing nickel thin film \cite{kim2017magnetization} based on the decomposition of magnetic and acoustic dynamics in phonon and magnon eigenmodes, respectively. The time-domain fitting of transient reflectivity data on both sides of the nickel film provides frequencies, lifetimes and phases of individual acoustic eigenmodes. The latter is shown to drive the magnetization dynamics, to be in quantitative agreement with time-resolved MOKE measurements. Notably, the comparison of MOKE signals on both sides of the sample evidence the in-phase FMR dynamics ($n=0$, with minor contributions of symmetric magnon eigenmodes $n=2,4,...$) induced by the lowest order (p=1) symmetric acoustic mode and the opposite-sign magnetization oscillation of antisymmetric magnon modes ($n=1,3,...$). Being in a quantitative agreement with a simple theoretical model with tabulated material parameters, the experimental data clearly evidence the resonantly enhanced excitation of nonuniform magnon modes. Moreover, accurate fitting of the magnetization dynamics driven by long-lived $p=1$ (9.8~GHz) and $p=2$ (19.1~GHz) acoustic modes delivers the correct value for magnetic Gilbert damping $\alpha=0.04$, corresponding to the quality factor $Q_m=13$ for magnon modes. Being smaller than the quality factors of acoustic modes, this magnetic quality factor assures optimum conditions for resonant phonon-magnon excitation, the phenomenon to be further explored in the ultrahigh THz-frequency regime \cite{vernik2022resonant,kimel20222022}.         

\begin{acknowledgments}
This article is dedicated to the memory of Jean-Yves Bigot, in whose labs the reported measurements have been performed. The support of the Physics Department of \'Ecole Polytechnique and Institut Polytechnique de Paris within the framework of a {\it Projet de Recherche en Laboratoire} and by the ANR-21-MRS1-0015-01 "IRON-MAG" is gratefully acknowledged. P.V. acknowledges support from the Spanish Ministry of Science and Innovation and the European Union under the Maria de Maeztu Units of Excellence Programme (CEX2020-001038-M) and the project PID2021-123943NB-I00 (MICINN/FEDER). J.K. acknowledges support from Basic Science Research Program through the National Research Foundation of Korea (NRF) funded by the Ministry of Education (2022R1I1A3072023) and by the MSIT (2021R1A4A1031920).
\end{acknowledgments}

%\bibliographystyle{unsrt}
%\bibliography{Biblio}

\begin{thebibliography}{28}%
\makeatletter
\providecommand \@ifxundefined [1]{%
 \@ifx{#1\undefined}
}%
\providecommand \@ifnum [1]{%
 \ifnum #1\expandafter \@firstoftwo
 \else \expandafter \@secondoftwo
 \fi
}%
\providecommand \@ifx [1]{%
 \ifx #1\expandafter \@firstoftwo
 \else \expandafter \@secondoftwo
 \fi
}%
\providecommand \natexlab [1]{#1}%
\providecommand \enquote  [1]{``#1''}%
\providecommand \bibnamefont  [1]{#1}%
\providecommand \bibfnamefont [1]{#1}%
\providecommand \citenamefont [1]{#1}%
\providecommand \href@noop [0]{\@secondoftwo}%
\providecommand \href [0]{\begingroup \@sanitize@url \@href}%
\providecommand \@href[1]{\@@startlink{#1}\@@href}%
\providecommand \@@href[1]{\endgroup#1\@@endlink}%
\providecommand \@sanitize@url [0]{\catcode `\\12\catcode `\$12\catcode
  `\&12\catcode `\#12\catcode `\^12\catcode `\_12\catcode `\%12\relax}%
\providecommand \@@startlink[1]{}%
\providecommand \@@endlink[0]{}%
\providecommand \url  [0]{\begingroup\@sanitize@url \@url }%
\providecommand \@url [1]{\endgroup\@href {#1}{\urlprefix }}%
\providecommand \urlprefix  [0]{URL }%
\providecommand \Eprint [0]{\href }%
\providecommand \doibase [0]{http://dx.doi.org/}%
\providecommand \selectlanguage [0]{\@gobble}%
\providecommand \bibinfo  [0]{\@secondoftwo}%
\providecommand \bibfield  [0]{\@secondoftwo}%
\providecommand \translation [1]{[#1]}%
\providecommand \BibitemOpen [0]{}%
\providecommand \bibitemStop [0]{}%
\providecommand \bibitemNoStop [0]{.\EOS\space}%
\providecommand \EOS [0]{\spacefactor3000\relax}%
\providecommand \BibitemShut  [1]{\csname bibitem#1\endcsname}%
\let\auto@bib@innerbib\@empty
%</preamble>
\bibitem [{\citenamefont {Kim}\ and\ \citenamefont
  {Bigot}(2017)}]{kim2017magnetization}%
  \BibitemOpen
  \bibfield  {author} {\bibinfo {author} {\bibfnamefont {J.-W.}\ \bibnamefont
  {Kim}}\ and\ \bibinfo {author} {\bibfnamefont {J.-Y.}\ \bibnamefont
  {Bigot}},\ }\href {\doibase 10.1103/PhysRevB.95.144422} {\bibfield  {journal}
  {\bibinfo  {journal} {Phys. Rev. B}\ }\textbf {\bibinfo {volume} {95}},\
  \bibinfo {pages} {144422} (\bibinfo {year} {2017})}\BibitemShut {NoStop}%
\bibitem [{\citenamefont {Vernik}\ \emph {et~al.}(2022)\citenamefont {Vernik},
  \citenamefont {Lomonosov}, \citenamefont {Vlasov}, \citenamefont {Kotov},
  \citenamefont {Kuzmin}, \citenamefont {Bychkov}, \citenamefont {Vavassori},\
  and\ \citenamefont {Temnov}}]{vernik2022resonant}%
  \BibitemOpen
  \bibfield  {author} {\bibinfo {author} {\bibfnamefont {U.}~\bibnamefont
  {Vernik}}, \bibinfo {author} {\bibfnamefont {A.~M.}\ \bibnamefont
  {Lomonosov}}, \bibinfo {author} {\bibfnamefont {V.~S.}\ \bibnamefont
  {Vlasov}}, \bibinfo {author} {\bibfnamefont {L.~N.}\ \bibnamefont {Kotov}},
  \bibinfo {author} {\bibfnamefont {D.~A.}\ \bibnamefont {Kuzmin}}, \bibinfo
  {author} {\bibfnamefont {I.~V.}\ \bibnamefont {Bychkov}}, \bibinfo {author}
  {\bibfnamefont {P.}~\bibnamefont {Vavassori}}, \ and\ \bibinfo {author}
  {\bibfnamefont {V.~V.}\ \bibnamefont {Temnov}},\ }\href@noop {} {\bibfield
  {journal} {\bibinfo  {journal} {Physical Review B}\ }\textbf {\bibinfo
  {volume} {106}},\ \bibinfo {pages} {144420} (\bibinfo {year}
  {2022})}\BibitemShut {NoStop}%
\bibitem [{\citenamefont {Scherbakov}\ \emph {et~al.}(2010)\citenamefont
  {Scherbakov}, \citenamefont {Salasyuk}, \citenamefont {Akimov}, \citenamefont
  {Liu}, \citenamefont {Bombeck}, \citenamefont {Br\"uggemann}, \citenamefont
  {Yakovlev}, \citenamefont {Sapega}, \citenamefont {Furdyna},\ and\
  \citenamefont {Bayer}}]{Scherbakov2010}%
  \BibitemOpen
  \bibfield  {author} {\bibinfo {author} {\bibfnamefont {A.~V.}\ \bibnamefont
  {Scherbakov}}, \bibinfo {author} {\bibfnamefont {A.~S.}\ \bibnamefont
  {Salasyuk}}, \bibinfo {author} {\bibfnamefont {A.~V.}\ \bibnamefont
  {Akimov}}, \bibinfo {author} {\bibfnamefont {X.}~\bibnamefont {Liu}},
  \bibinfo {author} {\bibfnamefont {M.}~\bibnamefont {Bombeck}}, \bibinfo
  {author} {\bibfnamefont {C.}~\bibnamefont {Br\"uggemann}}, \bibinfo {author}
  {\bibfnamefont {D.~R.}\ \bibnamefont {Yakovlev}}, \bibinfo {author}
  {\bibfnamefont {V.~F.}\ \bibnamefont {Sapega}}, \bibinfo {author}
  {\bibfnamefont {J.~K.}\ \bibnamefont {Furdyna}}, \ and\ \bibinfo {author}
  {\bibfnamefont {M.}~\bibnamefont {Bayer}},\ }\href {\doibase
  10.1103/PhysRevLett.105.117204} {\bibfield  {journal} {\bibinfo  {journal}
  {Phys. Rev. Lett.}\ }\textbf {\bibinfo {volume} {105}},\ \bibinfo {pages}
  {117204} (\bibinfo {year} {2010})}\BibitemShut {NoStop}%
\bibitem [{\citenamefont {Kim}\ \emph {et~al.}(2012)\citenamefont {Kim},
  \citenamefont {Vomir},\ and\ \citenamefont {Bigot}}]{kim2012ultrafast}%
  \BibitemOpen
  \bibfield  {author} {\bibinfo {author} {\bibfnamefont {J.-W.}\ \bibnamefont
  {Kim}}, \bibinfo {author} {\bibfnamefont {M.}~\bibnamefont {Vomir}}, \ and\
  \bibinfo {author} {\bibfnamefont {J.-Y.}\ \bibnamefont {Bigot}},\ }\href
  {\doibase 10.1103/PhysRevLett.109.166601} {\bibfield  {journal} {\bibinfo
  {journal} {Phys. Rev. Lett.}\ }\textbf {\bibinfo {volume} {109}},\ \bibinfo
  {pages} {166601} (\bibinfo {year} {2012})}\BibitemShut {NoStop}%
\bibitem [{\citenamefont {Thevenard}\ \emph {et~al.}(2010)\citenamefont
  {Thevenard}, \citenamefont {Peronne}, \citenamefont {Gourdon}, \citenamefont
  {Testelin}, \citenamefont {Cubukcu}, \citenamefont {Charron}, \citenamefont
  {Vincent}, \citenamefont {Lema\^{\i}tre},\ and\ \citenamefont
  {Perrin}}]{Thevenard2010}%
  \BibitemOpen
  \bibfield  {author} {\bibinfo {author} {\bibfnamefont {L.}~\bibnamefont
  {Thevenard}}, \bibinfo {author} {\bibfnamefont {E.}~\bibnamefont {Peronne}},
  \bibinfo {author} {\bibfnamefont {C.}~\bibnamefont {Gourdon}}, \bibinfo
  {author} {\bibfnamefont {C.}~\bibnamefont {Testelin}}, \bibinfo {author}
  {\bibfnamefont {M.}~\bibnamefont {Cubukcu}}, \bibinfo {author} {\bibfnamefont
  {E.}~\bibnamefont {Charron}}, \bibinfo {author} {\bibfnamefont
  {S.}~\bibnamefont {Vincent}}, \bibinfo {author} {\bibfnamefont
  {A.}~\bibnamefont {Lema\^{\i}tre}}, \ and\ \bibinfo {author} {\bibfnamefont
  {B.}~\bibnamefont {Perrin}},\ }\href {\doibase 10.1103/PhysRevB.82.104422}
  {\bibfield  {journal} {\bibinfo  {journal} {Phys. Rev. B}\ }\textbf {\bibinfo
  {volume} {82}},\ \bibinfo {pages} {104422} (\bibinfo {year}
  {2010})}\BibitemShut {NoStop}%
\bibitem [{\citenamefont {Kim}\ \emph {et~al.}(2015)\citenamefont {Kim},
  \citenamefont {Vomir},\ and\ \citenamefont {Bigot}}]{kim2015controlling}%
  \BibitemOpen
  \bibfield  {author} {\bibinfo {author} {\bibfnamefont {J.-W.}\ \bibnamefont
  {Kim}}, \bibinfo {author} {\bibfnamefont {M.}~\bibnamefont {Vomir}}, \ and\
  \bibinfo {author} {\bibfnamefont {J.-Y.}\ \bibnamefont {Bigot}},\ }\href
  {\doibase 10.1038/srep08511} {\bibfield  {journal} {\bibinfo  {journal}
  {Scientific reports}\ }\textbf {\bibinfo {volume} {5}},\ \bibinfo {pages}
  {8511} (\bibinfo {year} {2015})}\BibitemShut {NoStop}%
\bibitem [{\citenamefont {Kovalenko}\ \emph {et~al.}(2013)\citenamefont
  {Kovalenko}, \citenamefont {Pezeril},\ and\ \citenamefont
  {Temnov}}]{KovalenkoPRL2013}%
  \BibitemOpen
  \bibfield  {author} {\bibinfo {author} {\bibfnamefont {O.}~\bibnamefont
  {Kovalenko}}, \bibinfo {author} {\bibfnamefont {T.}~\bibnamefont {Pezeril}},
  \ and\ \bibinfo {author} {\bibfnamefont {V.~V.}\ \bibnamefont {Temnov}},\
  }\href {\doibase 10.1103/PhysRevLett.110.266602} {\bibfield  {journal}
  {\bibinfo  {journal} {Phys. Rev. Lett.}\ }\textbf {\bibinfo {volume} {110}},\
  \bibinfo {pages} {266602} (\bibinfo {year} {2013})}\BibitemShut {NoStop}%
\bibitem [{\citenamefont {Vlasov}\ \emph {et~al.}(2020)\citenamefont {Vlasov},
  \citenamefont {Lomonosov}, \citenamefont {Golov}, \citenamefont {Kotov},
  \citenamefont {Besse}, \citenamefont {Alekhin}, \citenamefont {Kuzmin},
  \citenamefont {Bychkov},\ and\ \citenamefont {Temnov}}]{Vlasov2020}%
  \BibitemOpen
  \bibfield  {author} {\bibinfo {author} {\bibfnamefont {V.~S.}\ \bibnamefont
  {Vlasov}}, \bibinfo {author} {\bibfnamefont {A.~M.}\ \bibnamefont
  {Lomonosov}}, \bibinfo {author} {\bibfnamefont {A.~V.}\ \bibnamefont
  {Golov}}, \bibinfo {author} {\bibfnamefont {L.~N.}\ \bibnamefont {Kotov}},
  \bibinfo {author} {\bibfnamefont {V.}~\bibnamefont {Besse}}, \bibinfo
  {author} {\bibfnamefont {A.}~\bibnamefont {Alekhin}}, \bibinfo {author}
  {\bibfnamefont {D.~A.}\ \bibnamefont {Kuzmin}}, \bibinfo {author}
  {\bibfnamefont {I.~V.}\ \bibnamefont {Bychkov}}, \ and\ \bibinfo {author}
  {\bibfnamefont {V.~V.}\ \bibnamefont {Temnov}},\ }\href {\doibase
  10.1103/PhysRevB.101.024425} {\bibfield  {journal} {\bibinfo  {journal}
  {Phys. Rev. B}\ }\textbf {\bibinfo {volume} {101}},\ \bibinfo {pages}
  {024425} (\bibinfo {year} {2020})}\BibitemShut {NoStop}%
\bibitem [{\citenamefont {Janu{\v{s}}onis}\ \emph
  {et~al.}(2016{\natexlab{a}})\citenamefont {Janu{\v{s}}onis}, \citenamefont
  {Chang}, \citenamefont {Jansma}, \citenamefont {Gatilova}, \citenamefont
  {Vlasov}, \citenamefont {Lomonosov}, \citenamefont {Temnov},\ and\
  \citenamefont {Tobey}}]{janusonis2016ultrafast}%
  \BibitemOpen
  \bibfield  {author} {\bibinfo {author} {\bibfnamefont {J.}~\bibnamefont
  {Janu{\v{s}}onis}}, \bibinfo {author} {\bibfnamefont {C.~L.}\ \bibnamefont
  {Chang}}, \bibinfo {author} {\bibfnamefont {T.}~\bibnamefont {Jansma}},
  \bibinfo {author} {\bibfnamefont {A.}~\bibnamefont {Gatilova}}, \bibinfo
  {author} {\bibfnamefont {V.~S.}\ \bibnamefont {Vlasov}}, \bibinfo {author}
  {\bibfnamefont {A.~M.}\ \bibnamefont {Lomonosov}}, \bibinfo {author}
  {\bibfnamefont {V.~V.}\ \bibnamefont {Temnov}}, \ and\ \bibinfo {author}
  {\bibfnamefont {R.~I.}\ \bibnamefont {Tobey}},\ }\href@noop {} {\bibfield
  {journal} {\bibinfo  {journal} {Phys. Rev. B}\ }\textbf {\bibinfo {volume}
  {94}},\ \bibinfo {pages} {024415} (\bibinfo {year}
  {2016}{\natexlab{a}})}\BibitemShut {NoStop}%
\bibitem [{\citenamefont {Janu{\v{s}}onis}\ \emph
  {et~al.}(2016{\natexlab{b}})\citenamefont {Janu{\v{s}}onis}, \citenamefont
  {Jansma}, \citenamefont {Chang}, \citenamefont {Q.}, \citenamefont
  {Gatilova}, \citenamefont {Lomonosov}, \citenamefont {Shalagatskyi},
  \citenamefont {Pezeril}, \citenamefont {Temnov},\ and\ \citenamefont
  {Tobey}}]{Janusonis2016_1}%
  \BibitemOpen
  \bibfield  {author} {\bibinfo {author} {\bibfnamefont {J.}~\bibnamefont
  {Janu{\v{s}}onis}}, \bibinfo {author} {\bibfnamefont {T.}~\bibnamefont
  {Jansma}}, \bibinfo {author} {\bibfnamefont {C.~L.}\ \bibnamefont {Chang}},
  \bibinfo {author} {\bibfnamefont {L.}~\bibnamefont {Q.}}, \bibinfo {author}
  {\bibfnamefont {A.}~\bibnamefont {Gatilova}}, \bibinfo {author}
  {\bibfnamefont {A.~M.}\ \bibnamefont {Lomonosov}}, \bibinfo {author}
  {\bibfnamefont {V.}~\bibnamefont {Shalagatskyi}}, \bibinfo {author}
  {\bibfnamefont {T.}~\bibnamefont {Pezeril}}, \bibinfo {author} {\bibfnamefont
  {V.~V.}\ \bibnamefont {Temnov}}, \ and\ \bibinfo {author} {\bibfnamefont
  {R.~I.}\ \bibnamefont {Tobey}},\ }\href@noop {} {\bibfield  {journal}
  {\bibinfo  {journal} {Scientific Reports}\ }\textbf {\bibinfo {volume} {6}},\
  \bibinfo {pages} {29143} (\bibinfo {year} {2016}{\natexlab{b}})}\BibitemShut
  {NoStop}%
\bibitem [{\citenamefont {Chang}\ \emph {et~al.}(2017)\citenamefont {Chang},
  \citenamefont {Lomonosov}, \citenamefont {Janusonis}, \citenamefont {Vlasov},
  \citenamefont {Temnov},\ and\ \citenamefont {Tobey}}]{chang2017parametric}%
  \BibitemOpen
  \bibfield  {author} {\bibinfo {author} {\bibfnamefont {C.~L.}\ \bibnamefont
  {Chang}}, \bibinfo {author} {\bibfnamefont {A.~M.}\ \bibnamefont
  {Lomonosov}}, \bibinfo {author} {\bibfnamefont {J.}~\bibnamefont
  {Janusonis}}, \bibinfo {author} {\bibfnamefont {V.~S.}\ \bibnamefont
  {Vlasov}}, \bibinfo {author} {\bibfnamefont {V.~V.}\ \bibnamefont {Temnov}},
  \ and\ \bibinfo {author} {\bibfnamefont {R.~I.}\ \bibnamefont {Tobey}},\
  }\href@noop {} {\bibfield  {journal} {\bibinfo  {journal} {Phys. Rev. B}\
  }\textbf {\bibinfo {volume} {95}},\ \bibinfo {pages} {060409(R)} (\bibinfo
  {year} {2017})}\BibitemShut {NoStop}%
\bibitem [{\citenamefont {Chang}\ \emph {et~al.}(2018)\citenamefont {Chang},
  \citenamefont {Mieszczak}, \citenamefont {Zelent}, \citenamefont {Besse},
  \citenamefont {Martens}, \citenamefont {Tamming}, \citenamefont {Janusonis},
  \citenamefont {Graczyk}, \citenamefont {M{\"u}nzenberg}, \citenamefont
  {K{\l}os} \emph {et~al.}}]{chang2018driving}%
  \BibitemOpen
  \bibfield  {author} {\bibinfo {author} {\bibfnamefont {C.~L.}\ \bibnamefont
  {Chang}}, \bibinfo {author} {\bibfnamefont {S.}~\bibnamefont {Mieszczak}},
  \bibinfo {author} {\bibfnamefont {M.}~\bibnamefont {Zelent}}, \bibinfo
  {author} {\bibfnamefont {V.}~\bibnamefont {Besse}}, \bibinfo {author}
  {\bibfnamefont {U.}~\bibnamefont {Martens}}, \bibinfo {author} {\bibfnamefont
  {R.~R.}\ \bibnamefont {Tamming}}, \bibinfo {author} {\bibfnamefont
  {J.}~\bibnamefont {Janusonis}}, \bibinfo {author} {\bibfnamefont
  {P.}~\bibnamefont {Graczyk}}, \bibinfo {author} {\bibfnamefont
  {M.}~\bibnamefont {M{\"u}nzenberg}}, \bibinfo {author} {\bibfnamefont
  {J.~W.}\ \bibnamefont {K{\l}os}},  \emph {et~al.},\ }\href@noop {} {\bibfield
   {journal} {\bibinfo  {journal} {Physical Review Applied}\ }\textbf {\bibinfo
  {volume} {10}},\ \bibinfo {pages} {064051} (\bibinfo {year}
  {2018})}\BibitemShut {NoStop}%
\bibitem [{\citenamefont {Van~Kampen}\ \emph {et~al.}(2002)\citenamefont
  {Van~Kampen}, \citenamefont {Jozsa}, \citenamefont {Kohlhepp}, \citenamefont
  {LeClair}, \citenamefont {Lagae}, \citenamefont {De~Jonge},\ and\
  \citenamefont {Koopmans}}]{van2002all}%
  \BibitemOpen
  \bibfield  {author} {\bibinfo {author} {\bibfnamefont {M.}~\bibnamefont
  {Van~Kampen}}, \bibinfo {author} {\bibfnamefont {C.}~\bibnamefont {Jozsa}},
  \bibinfo {author} {\bibfnamefont {J.}~\bibnamefont {Kohlhepp}}, \bibinfo
  {author} {\bibfnamefont {P.}~\bibnamefont {LeClair}}, \bibinfo {author}
  {\bibfnamefont {L.}~\bibnamefont {Lagae}}, \bibinfo {author} {\bibfnamefont
  {W.}~\bibnamefont {De~Jonge}}, \ and\ \bibinfo {author} {\bibfnamefont
  {B.}~\bibnamefont {Koopmans}},\ }\href {\doibase
  10.1103/PhysRevLett.88.227201} {\bibfield  {journal} {\bibinfo  {journal}
  {Phys. Rev. Lett.}\ }\textbf {\bibinfo {volume} {88}},\ \bibinfo {pages}
  {227201} (\bibinfo {year} {2002})}\BibitemShut {NoStop}%
\bibitem [{\citenamefont {Anisimov}\ \emph {et~al.}(1974)\citenamefont
  {Anisimov}, \citenamefont {Kapeliovich}, \citenamefont {Perelman} \emph
  {et~al.}}]{anisimov1974electron}%
  \BibitemOpen
  \bibfield  {author} {\bibinfo {author} {\bibfnamefont {S.}~\bibnamefont
  {Anisimov}}, \bibinfo {author} {\bibfnamefont {B.}~\bibnamefont
  {Kapeliovich}}, \bibinfo {author} {\bibfnamefont {T.}~\bibnamefont
  {Perelman}},  \emph {et~al.},\ }\href@noop {} {\bibfield  {journal} {\bibinfo
   {journal} {Zh. Eksp. Teor. Fiz}\ }\textbf {\bibinfo {volume} {66}},\
  \bibinfo {pages} {375} (\bibinfo {year} {1974})}\BibitemShut {NoStop}%
\bibitem [{\citenamefont {Beaurepaire}\ \emph {et~al.}(1996)\citenamefont
  {Beaurepaire}, \citenamefont {Merle}, \citenamefont {Daunois},\ and\
  \citenamefont {Bigot}}]{beaurepaire1996ultrafast}%
  \BibitemOpen
  \bibfield  {author} {\bibinfo {author} {\bibfnamefont {E.}~\bibnamefont
  {Beaurepaire}}, \bibinfo {author} {\bibfnamefont {J.-C.}\ \bibnamefont
  {Merle}}, \bibinfo {author} {\bibfnamefont {A.}~\bibnamefont {Daunois}}, \
  and\ \bibinfo {author} {\bibfnamefont {J.-Y.}\ \bibnamefont {Bigot}},\ }\href
  {\doibase 10.1103/PhysRevLett.76.4250} {\bibfield  {journal} {\bibinfo
  {journal} {Phys. Rev. Lett.}\ }\textbf {\bibinfo {volume} {76}},\ \bibinfo
  {pages} {4250} (\bibinfo {year} {1996})}\BibitemShut {NoStop}%
\bibitem [{\citenamefont {G{\"u}dde}\ \emph {et~al.}(1999)\citenamefont
  {G{\"u}dde}, \citenamefont {Conrad}, \citenamefont {J{\"a}hnke},
  \citenamefont {Hohlfeld},\ and\ \citenamefont
  {Matthias}}]{gudde1999magnetization}%
  \BibitemOpen
  \bibfield  {author} {\bibinfo {author} {\bibfnamefont {J.}~\bibnamefont
  {G{\"u}dde}}, \bibinfo {author} {\bibfnamefont {U.}~\bibnamefont {Conrad}},
  \bibinfo {author} {\bibfnamefont {V.}~\bibnamefont {J{\"a}hnke}}, \bibinfo
  {author} {\bibfnamefont {J.}~\bibnamefont {Hohlfeld}}, \ and\ \bibinfo
  {author} {\bibfnamefont {E.}~\bibnamefont {Matthias}},\ }\href@noop {}
  {\bibfield  {journal} {\bibinfo  {journal} {Physical Review B}\ }\textbf
  {\bibinfo {volume} {59}},\ \bibinfo {pages} {R6608} (\bibinfo {year}
  {1999})}\BibitemShut {NoStop}%
\bibitem [{\citenamefont {Saito}\ \emph {et~al.}(2003)\citenamefont {Saito},
  \citenamefont {Matsuda},\ and\ \citenamefont {Wright}}]{saito2003picosecond}%
  \BibitemOpen
  \bibfield  {author} {\bibinfo {author} {\bibfnamefont {T.}~\bibnamefont
  {Saito}}, \bibinfo {author} {\bibfnamefont {O.}~\bibnamefont {Matsuda}}, \
  and\ \bibinfo {author} {\bibfnamefont {O.}~\bibnamefont {Wright}},\
  }\href@noop {} {\bibfield  {journal} {\bibinfo  {journal} {Physical Review
  B}\ }\textbf {\bibinfo {volume} {67}},\ \bibinfo {pages} {205421} (\bibinfo
  {year} {2003})}\BibitemShut {NoStop}%
\bibitem [{\citenamefont {Chikazumi}\ and\ \citenamefont
  {Graham}(1997)}]{chikazumi1997physics}%
  \BibitemOpen
  \bibfield  {author} {\bibinfo {author} {\bibfnamefont {S.}~\bibnamefont
  {Chikazumi}}\ and\ \bibinfo {author} {\bibfnamefont {C.~D.}\ \bibnamefont
  {Graham}},\ }\href@noop {} {\emph {\bibinfo {title} {Physics of
  ferromagnetism}}},\ \bibinfo {number} {94}\ (\bibinfo  {publisher} {Oxford
  university press},\ \bibinfo {year} {1997})\BibitemShut {NoStop}%
\bibitem [{\citenamefont {Besse}\ \emph {et~al.}(2020)\citenamefont {Besse},
  \citenamefont {Golov}, \citenamefont {Vlasov}, \citenamefont {Alekhin},
  \citenamefont {Kuzmin}, \citenamefont {Bychkov}, \citenamefont {Kotov},\ and\
  \citenamefont {Temnov}}]{BesseJMMM}%
  \BibitemOpen
  \bibfield  {author} {\bibinfo {author} {\bibfnamefont {V.}~\bibnamefont
  {Besse}}, \bibinfo {author} {\bibfnamefont {A.~V.}\ \bibnamefont {Golov}},
  \bibinfo {author} {\bibfnamefont {V.~S.}\ \bibnamefont {Vlasov}}, \bibinfo
  {author} {\bibfnamefont {A.}~\bibnamefont {Alekhin}}, \bibinfo {author}
  {\bibfnamefont {D.}~\bibnamefont {Kuzmin}}, \bibinfo {author} {\bibfnamefont
  {I.~V.}\ \bibnamefont {Bychkov}}, \bibinfo {author} {\bibfnamefont {L.~N.}\
  \bibnamefont {Kotov}}, \ and\ \bibinfo {author} {\bibfnamefont {V.~V.}\
  \bibnamefont {Temnov}},\ }\href@noop {} {\bibfield  {journal} {\bibinfo
  {journal} {J. Magn. Magn. Mater.}\ }\textbf {\bibinfo {volume} {502}},\
  \bibinfo {pages} {166320} (\bibinfo {year} {2020})}\BibitemShut {NoStop}%
\bibitem [{\citenamefont {Thomsen}\ \emph {et~al.}(1986)\citenamefont
  {Thomsen}, \citenamefont {Grahn}, \citenamefont {Maris},\ and\ \citenamefont
  {Tauc}}]{ThomsenPRB86}%
  \BibitemOpen
  \bibfield  {author} {\bibinfo {author} {\bibfnamefont {C.}~\bibnamefont
  {Thomsen}}, \bibinfo {author} {\bibfnamefont {H.~T.}\ \bibnamefont {Grahn}},
  \bibinfo {author} {\bibfnamefont {H.~J.}\ \bibnamefont {Maris}}, \ and\
  \bibinfo {author} {\bibfnamefont {J.}~\bibnamefont {Tauc}},\ }\href {\doibase
  10.1103/PhysRevB.34.4129} {\bibfield  {journal} {\bibinfo  {journal} {Phys.
  Rev. B}\ }\textbf {\bibinfo {volume} {34}},\ \bibinfo {pages} {4129}
  (\bibinfo {year} {1986})}\BibitemShut {NoStop}%
\bibitem [{\citenamefont {Zhu}\ \emph {et~al.}(1991)\citenamefont {Zhu},
  \citenamefont {Maris},\ and\ \citenamefont {Tauc}}]{Maris}%
  \BibitemOpen
  \bibfield  {author} {\bibinfo {author} {\bibfnamefont {T.~C.}\ \bibnamefont
  {Zhu}}, \bibinfo {author} {\bibfnamefont {H.~J.}\ \bibnamefont {Maris}}, \
  and\ \bibinfo {author} {\bibfnamefont {J.}~\bibnamefont {Tauc}},\ }\href
  {\doibase 10.1103/PhysRevB.44.4281} {\bibfield  {journal} {\bibinfo
  {journal} {Phys. Rev. B}\ }\textbf {\bibinfo {volume} {44}},\ \bibinfo
  {pages} {4281} (\bibinfo {year} {1991})}\BibitemShut {NoStop}%
\bibitem [{\citenamefont {Lin}\ \emph {et~al.}(2008)\citenamefont {Lin},
  \citenamefont {Zhigilei},\ and\ \citenamefont {Celli}}]{lin2008electron}%
  \BibitemOpen
  \bibfield  {author} {\bibinfo {author} {\bibfnamefont {Z.}~\bibnamefont
  {Lin}}, \bibinfo {author} {\bibfnamefont {L.~V.}\ \bibnamefont {Zhigilei}}, \
  and\ \bibinfo {author} {\bibfnamefont {V.}~\bibnamefont {Celli}},\
  }\href@noop {} {\bibfield  {journal} {\bibinfo  {journal} {Physical Review
  B}\ }\textbf {\bibinfo {volume} {77}},\ \bibinfo {pages} {075133} (\bibinfo
  {year} {2008})}\BibitemShut {NoStop}%
\bibitem [{\citenamefont {Traeger}\ \emph {et~al.}(1992)\citenamefont
  {Traeger}, \citenamefont {Wenzel},\ and\ \citenamefont
  {Hubert}}]{traeger1992depthF}%
  \BibitemOpen
  \bibfield  {author} {\bibinfo {author} {\bibfnamefont {G.}~\bibnamefont
  {Traeger}}, \bibinfo {author} {\bibfnamefont {L.}~\bibnamefont {Wenzel}}, \
  and\ \bibinfo {author} {\bibfnamefont {A.}~\bibnamefont {Hubert}},\
  }\href@noop {} {\bibfield  {journal} {\bibinfo  {journal} {physica status
  solidi (a)}\ }\textbf {\bibinfo {volume} {131}},\ \bibinfo {pages} {201}
  (\bibinfo {year} {1992})}\BibitemShut {NoStop}%
\bibitem [{\citenamefont {Krinchik}\ and\ \citenamefont
  {Artemjev}(1968)}]{Krinchik1968}%
  \BibitemOpen
  \bibfield  {author} {\bibinfo {author} {\bibfnamefont {G.~S.}\ \bibnamefont
  {Krinchik}}\ and\ \bibinfo {author} {\bibfnamefont {V.~A.}\ \bibnamefont
  {Artemjev}},\ }\href {\doibase 10.1063/1.1656263} {\bibfield  {journal}
  {\bibinfo  {journal} {Journal of Applied Physics}\ }\textbf {\bibinfo
  {volume} {39}},\ \bibinfo {pages} {1276} (\bibinfo {year} {1968})},\ \Eprint
  {http://arxiv.org/abs/https://doi.org/10.1063/1.1656263}
  {https://doi.org/10.1063/1.1656263} \BibitemShut {NoStop}%
\bibitem [{\citenamefont {Berk}\ \emph {et~al.}(2019)\citenamefont {Berk},
  \citenamefont {Jaris}, \citenamefont {Yang}, \citenamefont {Dhuey},
  \citenamefont {Cabrini},\ and\ \citenamefont {Schmidt}}]{berk2019strongly}%
  \BibitemOpen
  \bibfield  {author} {\bibinfo {author} {\bibfnamefont {C.}~\bibnamefont
  {Berk}}, \bibinfo {author} {\bibfnamefont {M.}~\bibnamefont {Jaris}},
  \bibinfo {author} {\bibfnamefont {W.}~\bibnamefont {Yang}}, \bibinfo {author}
  {\bibfnamefont {S.}~\bibnamefont {Dhuey}}, \bibinfo {author} {\bibfnamefont
  {S.}~\bibnamefont {Cabrini}}, \ and\ \bibinfo {author} {\bibfnamefont
  {H.}~\bibnamefont {Schmidt}},\ }\href@noop {} {\bibfield  {journal} {\bibinfo
   {journal} {Nature communications}\ }\textbf {\bibinfo {volume} {10}},\
  \bibinfo {pages} {1} (\bibinfo {year} {2019})}\BibitemShut {NoStop}%
\bibitem [{\citenamefont {Razdolski}\ \emph {et~al.}(2017)\citenamefont
  {Razdolski}, \citenamefont {Alekhin}, \citenamefont {Ilin}, \citenamefont
  {Meyburg}, \citenamefont {Roddatis}, \citenamefont {Diesing}, \citenamefont
  {Bovensiepen},\ and\ \citenamefont {Melnikov}}]{razdolski2017nanoscale}%
  \BibitemOpen
  \bibfield  {author} {\bibinfo {author} {\bibfnamefont {I.}~\bibnamefont
  {Razdolski}}, \bibinfo {author} {\bibfnamefont {A.}~\bibnamefont {Alekhin}},
  \bibinfo {author} {\bibfnamefont {N.}~\bibnamefont {Ilin}}, \bibinfo {author}
  {\bibfnamefont {J.~P.}\ \bibnamefont {Meyburg}}, \bibinfo {author}
  {\bibfnamefont {V.}~\bibnamefont {Roddatis}}, \bibinfo {author}
  {\bibfnamefont {D.}~\bibnamefont {Diesing}}, \bibinfo {author} {\bibfnamefont
  {U.}~\bibnamefont {Bovensiepen}}, \ and\ \bibinfo {author} {\bibfnamefont
  {A.}~\bibnamefont {Melnikov}},\ }\href@noop {} {\bibfield  {journal}
  {\bibinfo  {journal} {Nature communications}\ }\textbf {\bibinfo {volume}
  {8}},\ \bibinfo {pages} {15007} (\bibinfo {year} {2017})}\BibitemShut
  {NoStop}%
\bibitem [{\citenamefont {Salikhov}\ \emph {et~al.}(2019)\citenamefont
  {Salikhov}, \citenamefont {Alekhin}, \citenamefont {Parpiiev}, \citenamefont
  {Pezeril}, \citenamefont {Makarov}, \citenamefont {Abrudan}, \citenamefont
  {Meckenstock}, \citenamefont {Radu}, \citenamefont {Farle}, \citenamefont
  {Zabel},\ and\ \citenamefont {Temnov}}]{Salikhov2019}%
  \BibitemOpen
  \bibfield  {author} {\bibinfo {author} {\bibfnamefont {R.}~\bibnamefont
  {Salikhov}}, \bibinfo {author} {\bibfnamefont {A.}~\bibnamefont {Alekhin}},
  \bibinfo {author} {\bibfnamefont {T.}~\bibnamefont {Parpiiev}}, \bibinfo
  {author} {\bibfnamefont {T.}~\bibnamefont {Pezeril}}, \bibinfo {author}
  {\bibfnamefont {D.}~\bibnamefont {Makarov}}, \bibinfo {author} {\bibfnamefont
  {R.}~\bibnamefont {Abrudan}}, \bibinfo {author} {\bibfnamefont
  {R.}~\bibnamefont {Meckenstock}}, \bibinfo {author} {\bibfnamefont
  {F.}~\bibnamefont {Radu}}, \bibinfo {author} {\bibfnamefont {M.}~\bibnamefont
  {Farle}}, \bibinfo {author} {\bibfnamefont {H.}~\bibnamefont {Zabel}}, \ and\
  \bibinfo {author} {\bibfnamefont {V.~V.}\ \bibnamefont {Temnov}},\ }\href
  {\doibase 10.1103/PhysRevB.99.104412} {\bibfield  {journal} {\bibinfo
  {journal} {Phys. Rev. B}\ }\textbf {\bibinfo {volume} {99}},\ \bibinfo
  {pages} {104412} (\bibinfo {year} {2019})}\BibitemShut {NoStop}%
\bibitem [{\citenamefont {Kimel}\ \emph {et~al.}(2022)\citenamefont {Kimel},
  \citenamefont {Zvezdin}, \citenamefont {Sharma}, \citenamefont {Shallcross},
  \citenamefont {De~Sousa}, \citenamefont {Garc{\'\i}a-Mart{\'\i}n},
  \citenamefont {Salvan}, \citenamefont {Hamrle}, \citenamefont {Stejskal},
  \citenamefont {McCord} \emph {et~al.}}]{kimel20222022}%
  \BibitemOpen
  \bibfield  {author} {\bibinfo {author} {\bibfnamefont {A.}~\bibnamefont
  {Kimel}}, \bibinfo {author} {\bibfnamefont {A.}~\bibnamefont {Zvezdin}},
  \bibinfo {author} {\bibfnamefont {S.}~\bibnamefont {Sharma}}, \bibinfo
  {author} {\bibfnamefont {S.}~\bibnamefont {Shallcross}}, \bibinfo {author}
  {\bibfnamefont {N.}~\bibnamefont {De~Sousa}}, \bibinfo {author}
  {\bibfnamefont {A.}~\bibnamefont {Garc{\'\i}a-Mart{\'\i}n}}, \bibinfo
  {author} {\bibfnamefont {G.}~\bibnamefont {Salvan}}, \bibinfo {author}
  {\bibfnamefont {J.}~\bibnamefont {Hamrle}}, \bibinfo {author} {\bibfnamefont
  {O.}~\bibnamefont {Stejskal}}, \bibinfo {author} {\bibfnamefont
  {J.}~\bibnamefont {McCord}},  \emph {et~al.},\ }\href@noop {} {\bibfield
  {journal} {\bibinfo  {journal} {Journal of Physics D: Applied Physics}\
  }\textbf {\bibinfo {volume} {55}},\ \bibinfo {pages} {463003} (\bibinfo
  {year} {2022})}\BibitemShut {NoStop}%
\end{thebibliography}
%merlin.mbs apsrev4-1.bst 2010-07-25 4.21a (PWD, AO, DPC) hacked
%Control: key (0)
%Control: author (72) initials jnrlst
%Control: editor formatted (1) identically to author
%Control: production of article title (-1) disabled
%Control: page (0) single
%Control: year (1) truncated
%Control: production of eprint (0) enabled
%

\end{document}